%% file: manuscript.tex
\documentclass[sigconf]{acmart}

\usepackage{booktabs}
\usepackage{tabularx}
\usepackage{array}
\usepackage{longtable} 
\usepackage{multirow}
\usepackage{xcolor}

\newcolumntype{Y}{>{\raggedright\arraybackslash}X}
\AtBeginDocument{%
  }

\setcopyright{none}
\acmConference[under review HSCC ICCPS '27]{30th ACM/IEEE International Conference on Cyber-Physical and Autonomous Systems}{May 2027}{Boulder, CO, USA}
\renewcommand\footnotetextcopyrightpermission[1]{}

\begin{document}

\title{Toward Secure AI at the Edge \\Threats, Defenses, and Open Research Challenges}


\author{Sean Oesch}
\correspondingauthor
\affiliation{%
  \institution{Oak Ridge National Laboratory}
  \city{Oak Ridge}
  \state{TN}
  \country{USA}
}
\email{oeschts@ornl.gov}

\author{Jack Hutchins}
\affiliation{%
  \institution{Oak Ridge National Laboratory}
  \city{Oak Ridge}
  \state{TN}
  \country{USA}
}

\author{Samuel Hildebrand}
\affiliation{%
  \institution{Oak Ridge National Laboratory}
  \city{Oak Ridge}
  \state{TN}
  \country{USA}
}

\author{Brian Weber}
\affiliation{%
  \institution{Oak Ridge National Laboratory}
  \city{Oak Ridge}
  \state{TN}
  \country{USA}
}

\author{Luke Koch}
\correspondingauthor
\affiliation{%
  \institution{Oak Ridge National Laboratory}
  \city{Oak Ridge}
  \state{TN}
  \country{USA}}
\email{kochlr@ornl.gov}

\author{Edmon Begoli}
\affiliation{%
  \institution{Oak Ridge National Laboratory}
  \city{Oak Ridge}
  \state{TN}
  \country{USA}}
  \email{begolie@ornl.gov}


\renewcommand{\shortauthors}{Oesch et al.}

\newcommand{\shayan}[1]{\textcolor{red}{[SM: #1]}}

\definecolor{nonAiColor}{HTML}{1F4E79} 
\definecolor{aiColor}{HTML}{C55A11} 
\definecolor{envColor}{HTML}{2A9D8F} 

\begin{abstract}
AI at the edge is increasingly used to provide high speed, low latency analytics and decision making capabilities in critical applications such as autonomous systems.
Unlike traditional AI models deployed in well resourced
data centers and protected by multiple layers of security, AI models deployed on edge devices are more vulnerable to
tamper and theft.
In this paper we explore ways attackers exploit edge AI, how defenders can protect sensitive models, and lay out a vision for future research to enhance the security of AI models deployed at the edge.
\end{abstract}



\maketitle

\section{Introduction}
AI at the edge is increasingly used to provide high speed, low latency analytics and decision making capabilities in critical applications such as autonomous systems (drones, robotics), healthcare (e.g. glucose sensors, heartbeat monitors), agriculture, and the energy grid~\cite{chang2021survey}.
Unlike traditional AI models deployed in well resourced data centers and protected by multiple layers of security, AI models deployed on low resource edge devices are more vulnerable to tamper and theft. 
Given that the market for edge AI is predicted to grow from 312 billion in 2026 to 11.86 billion USD in 2032~\footnote{\url{https://finance.yahoo.com/technology/ai/articles/edge-ai-software-market-growth-153800461.html}} and that security of edge AI models is not keeping pace with the growth in their usage~\footnote{\url{https://www.dell.com/en-us/blog/secure-edge-ai-before-it-breaks/}}, it is imperative for the security community to understand the unique security landscape of edge AI and advocate for best practices to secure AI at the edge in high consequence applications.

AI security research traditionally focuses on attacks against training data or algorithm integrity such as data poisoning, model evasion, and model inference, but pays little attention to device security. 
Embedded security is concerned with protecting firmware, cryptographic keys, and communication protocols. 
Edge AI presents a novel attack surface because it sits at the intersection of AI security and embedded security. 

Edge AI security depends on interactions between the model, runtime, memory, processor, accelerator, firmware, and the physical device.
A neural network may be encrypted while stored and authenticated during boot, yet expose its parameters during runtime. 
A trusted execution environment may protect against an attacker with privileged access to the operating system and yet fail to protect against model theft through power analysis or other side-channel attacks. 
And an attacker with full control of the device may exploit debug interfaces, insecure firmware, or chip-modification to extract the model. 
 
Moreover, if the edge device sends and/or receives data from another network, the security of that network is assumed. 
In 2021 the security camera company Verkada failed to implement basic security measures such as unique and complex passwords, encryption of customer data, and proper network controls.
As a result, hackers gained access to 150,000 live customer cameras where AI was used for inference at the edge, as well as physical addresses, audio recordings, and customer WiFi credentials.~\footnote{\url{https://www.ftc.gov/news-events/news/press-releases/2024/08/ftc-takes-action-against-security-camera-firm-verkada-over-charges-it-failed-secure-videos-other}} 

This paper begins by providing a robust definition of Edge AI that includes subfields such as TinyML and elucidates the differences between these subfields. 
We then discuss attacks and defenses at the software and hardware levels, with an emphasis on helping the reader understand the costs associated with specific attacks and which defenses protect against which categories of attacks. 
In closing, we discuss open research questions and suggest a concrete path forward to secure AI at the edge. 

\section{Defining Edge AI}
\begin{figure*}[t]
    \centering
    \includegraphics[width=0.8\textwidth]{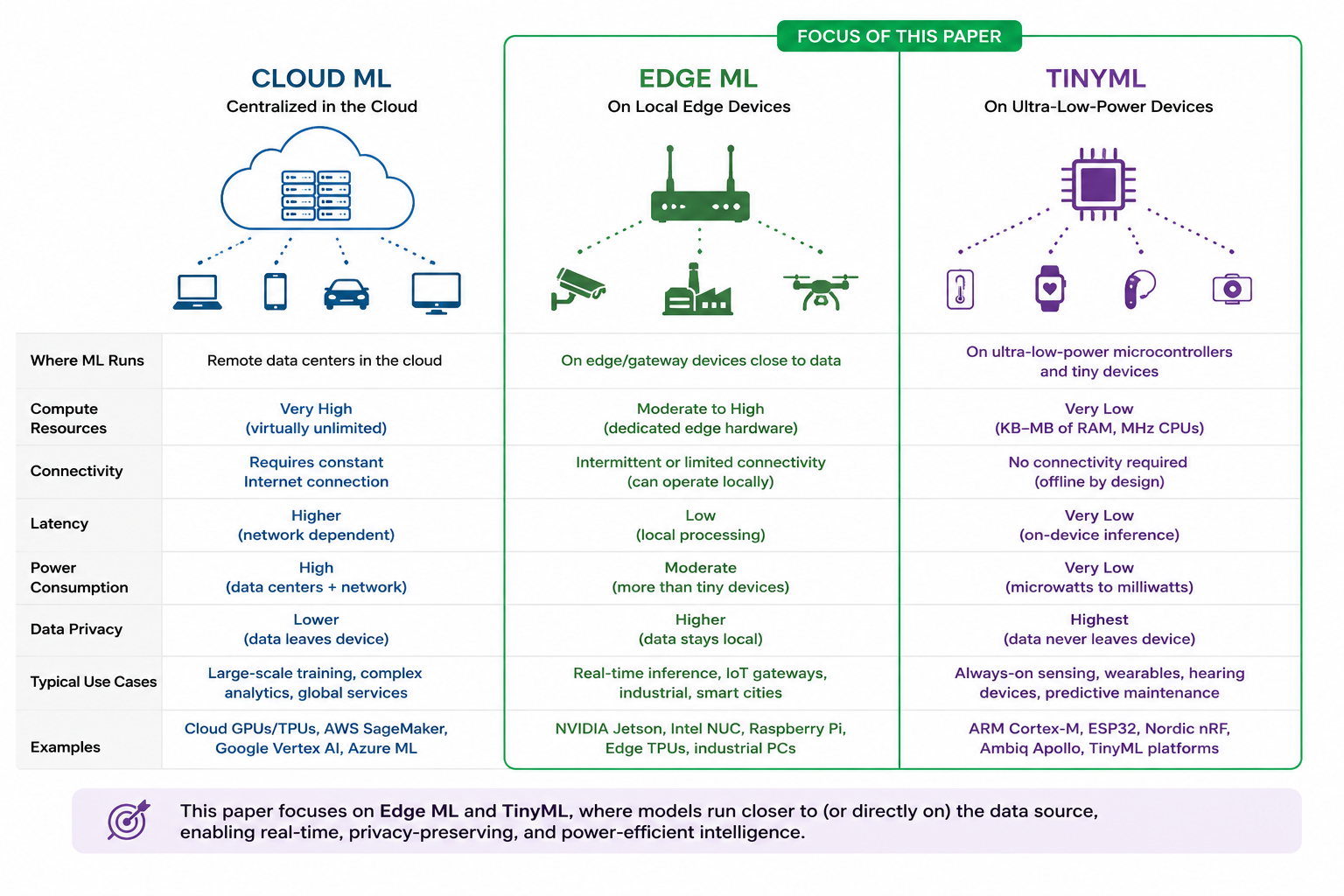}
    \caption{Comparison of Cloud ML, Edge ML, and TinyML. Edge ML and
    TinyML are the focus of this paper.}
    \label{fig:ml-deployment-comparison}
\end{figure*}

Deep learning models are often run in the cloud rather than at the edge to enable the use of large compute clusters.
Data is sent from the edge to the cloud, aggregated, analyzed, and then any relevant results are propagated back out to the edge. 
However, as shown in Figure~\ref{fig:ml-deployment-comparison}, Edge AI runs on local edge devices or in the case of TinyML on ultra-low-power devices rather than in the cloud. 

In Edge AI, data is collected and analyzed on the edge device where it is generated. 
Edge AI enables real-time decisions, enhances data privacy, reduces bandwidth requirements, and allows systems to run offline.
Applications of Edge AI range from sensors that detect when a component in a manufacturing pipeline is about to fail to wearable health monitors that send an alert when a life threatening anomaly is detected to decision making in autonomous drones. 

\section{Attacks Against Edge AI} 
Research has largely demonstrated that AI systems are vulnerable to unique attacks known collectively as counter-AI or, more specifically, adversarial machine learning methods \cite {goodfellow2014, koch2025adversarial}. These methods manipulate the input and/or extracted features fed into an AI model to degrade its performance. Examples of inference-time attacks include polyglot malware that prevents full feature extraction \cite {koch2025abuse} or adversarial images whose pixels have been optimized for misclassification \cite{kurakin2018adversarial}. Other categories of attack include data poisoning which degrades a model during training or membership inference which reveals specific data used to train the model \cite{luan2025survey}.

Edge deployment of AI adds an additional attack surface: hardware access. An attacker may gain physical access to the device and evaluate/extract the model using debug protocols, firmware extraction, or side-channel analysis. Firmware extraction can leverage physical fault injection methods like power glitching to bypass software-based defenses. Furthermore, advanced side-channel analysis methods leveraging radio frequency (RF) or other electromagnetic (EM) emissions can recover cryptographic keys or model weights when all other methods of gaining access or prevented. 

Once an adversary is able to extract the weights or benchmark the model in-situ, they can optimize adversarial ML attacks to defeat the same or similar models deployed elsewhere. The experimentally established transferability of adversarial attacks means that attacks against an extracted or surrogate model are likely to be effective against other models optimized for the same task despite architectural differences \cite{szegedy2014intriguingpropertiesneuralnetworks}. Extracted communication protocols and firmware also allow the adversary to craft attacks against the remaining devices deployed in the field. While traditional cyber attacks that take down the remaining deployed Edge AI devices are a significant concern, a more pernicious threat is that the attacker could deploy subtle AI model degradations that only trigger under specific conditions.

Edge AI deployment, therefore, makes protecting a model against extraction or attack optimization a significant challenge. Trusted execution environments are a promising example of how to secure Edge AI systems against some level of attack. However, against an adversary with significant side-channel analytic capabilities, we argue that further methods are vital. 

\input{defense.tex}

\section{Open Research Questions}

Edge devices often operate under tight resource constraints.
Common protection mechanisms such as tamper detection and side-channel countermeasures may impose unacceptable costs. 
As a result of these costs, securing AI at the edge requires a security-by-design approach where vulnerabilities are chosen based on a specific threat model rather than coincidental. 
An example can be seen in the work of Hernandez et al.~\cite{hernandez2026onnx2tee}, which partitions ONNX
models using graph-aware segmentation to run security sensitive portions of the model in a trusted execution environment while minimizing latency. 
More research is needed to maximize protection while minimizing resource usage and to develop best practices for security-by-design in EdgeAI systems. 

\subsection{Standardized Security Evaluation} Research on edge AI security lacks a consistent basis for comparing attacks and defenses across hardware platforms and attacker capabilities. Future work should establish reproducible evaluation methodologies that report model-recovery fidelity, attack success, query and trace requirements, attacker effort, and implementation overhead. Evaluations should consider adaptive adversaries and clearly distinguish resistance to extraction from resistance to functional replication or adversarial misuse. A common benchmark would enable more meaningful comparisons and help determine whether a defense provides practical security under a specified threat model.

\subsection{Resource-aware Defense Composition} Protection mechanisms such as masking, shuffling, trusted execution, and runtime transformations impose different costs in energy, latency, memory, and hardware area. A key challenge is determining which combinations provide the greatest security benefit within the resource budgets of a particular device. Research should investigate how to select and compose defenses based on attacker capabilities and application requirements, while measuring security and operational costs together. Such work should also establish whether the benefits of combining defenses persist against adaptive attacks, rather than assuming that individually effective mechanisms provide additive protection.

\subsection{Fleet-level Security under Partial Compromise} Per-device model diversity and hardware binding may limit the consequences of extracting a model from a single device, but their effectiveness may diminish against attackers who capture multiple devices. Future work should quantify how model diversity affects attack transferability and model recovery as the number of compromised devices increases. Evaluations should include the costs of generating, validating, provisioning, and updating device-specific models. More broadly, fleet defenses must account for shared software vulnerabilities, compromised signing credentials, and attacks that persist after individual devices have been revoked.

\subsection{Safe Containment and Recovery} Detecting a compromised device does not necessarily provide a safe response. Revocation and quarantine may prevent further misuse but can also disrupt safety-critical functions, particularly when devices operate offline or have limited communication. Research should develop recovery policies that balance false alarms, missed compromises, availability, and operational safety. Relevant questions include how devices should behave when attestation is unavailable, how trust can be re-established after compromise, and which capabilities should remain available during recovery. These policies should be evaluated under realistic failure and attack scenarios rather than solely by detection accuracy.

Addressing these questions requires a security-by-design approach that evaluates the complete edge AI system under explicit attacker capabilities and measurable resource constraints. Researchers should seek to establish which security properties can be achieved, at what cost, and  what residual risk is left over.

\bibliographystyle{ACM-Reference-Format}
\bibliography{references}


\end{document}

%% file: defense.tex
\section{Defending Edge AI}

Similar to in enterprise security, no single mechanism secures an edge AI device. For example, A trusted execution
environment (TEE) can isolate a model from untrusted software, but isolation
alone does not stop physical attacks or make the model robust to adversarial
inputs. It also does nothing for the rest of the fleet once one device falls.
Throughout, we argue that rather than judging defenses against a generic notion
of ``secure,'' it's possible to obtain a more objective measure of security by designing defenses against explicit threat
models, constructed using the following three questions: What can an attacker
obtain from a single device? What does that exposure enable against other
devices? And how can defenders detect and recover from unauthorized changes?

\subsection{Threat Model and Security Goals}

We consider four attacker profiles, ordered roughly by the access they
require:
\begin{description}
  \item A1 (input and query access): can choose inputs, such as sensor
  stimuli, and observe the outputs the device exposes by design.
  \item A2 (software control): controls an application, operating system,
  or driver outside the trusted boundary.
  \item A3 (device capture): holds a device and can read its storage, use
  debug interfaces, measure power or electromagnetic emissions, and attempt
  limited fault injection.
  \item A4 (laboratory access): can additionally perform invasive or
  semi-invasive procedures such as depackaging and microprobing.
\end{description}
These profiles are not a strict hierarchy. Some fault attacks can be
triggered from software (A2), and physically holding a device (A3) does not
automatically grant control of its trusted execution.

A meaningful security claim states the attacker's budget: how many queries
or side-channel traces, how much time and equipment, and how many captured
devices. We treat supply-chain compromise and theft of signing keys as
separate scenarios. Our main capture scenario assumes the model owner and
release process are trustworthy. A backdoor shipped through the official
release channel violates that assumption, and we discuss it under fleet
monitoring below.

Defenders typically need several distinct properties:
\begin{itemize}
  \item confidentiality of the model and its keys;
  \item integrity of the model, its configuration, and its updates;
  \item correct execution of inference from input to output;
  \item robustness to adversarial inputs and protection of training data;
  \item the ability to contain and recover from compromise across the fleet.
\end{itemize}
These goals are independent. Perfectly protected weights can still be fooled
by an adversarial image, and an officially released model can still contain
a backdoor. Table~\ref{tab:defense-map} summarizes which defenses address
which goals and where each falls short.

\begin{table*}[t]
\centering
\small
\caption{Defense roles under attacker profiles A1--A4. Actual protection
depends on the implementation; resisting A4 requires a separately justified
physical protection boundary.}
\label{tab:defense-map}
\renewcommand{\arraystretch}{1.08}
\begin{tabularx}{\textwidth}{@{}>{\raggedright\arraybackslash}p{0.21\textwidth}>{\raggedright\arraybackslash}p{0.14\textwidth}YY@{}}
\toprule
Defense & Profiles & Main contribution & Residual limitation \\
\midrule
Authenticated boot, updates, and rollback prevention & A2--A3 &
Only approved software and model versions run &
An approved artifact may itself be malicious.\\
\midrule
TEE isolation & A2; some A3 &
Protects code, data, keys, and provisioning from untrusted software &
Peripheral coverage and physical resistance vary by platform.\\
\midrule
FPGA configuration protection & A2--A3 &
Keeps the hardware design confidential and authentic &
External weights, buses, and runtime leakage need separate protection.\\
\midrule
Masking and hiding & Some A3 &
Raises the cost of side-channel attacks &
Costly; must be evaluated against adaptive attackers.\\
\midrule
Per-device model diversity & After extraction &
Limits how well an attack on one device transfers to others &
Does not help if devices share credentials or software flaws.\\
\midrule
Hardware-bound models & A1--A3, scheme-dependent &
A copied model does not run correctly elsewhere &
Query access or functional extraction may bypass it.\\
\midrule
Attestation and monitoring & A2--A3 &
Evidence that a device is in an approved state and behaving normally &
Cannot cover untested inputs or guarantee future behavior.\\
\midrule
Watermarks and fingerprints & After disclosure &
Proves ownership or identifies the leaking device &
Does not prevent private use of a stolen model.\\
\bottomrule
\end{tabularx}
\end{table*}

\subsection{Trusted Execution Environments}

A TEE runs sensitive code in an isolated region that the rest of the system,
including the operating system, cannot read or modify. For edge AI, its most
useful role is gatekeeping. A device first proves through remote attestation
that it is in an approved state, and only then receives the model or the key
to decrypt it. The IETF RATS architecture explicitly describes this kind of
attestation-gated provisioning \cite{birkholz2023rats}. How much protection a
TEE provides in practice depends on the specific processor, trusted runtime,
memory configuration, and which peripherals fall inside the boundary.

Edge TEEs usually have little secure memory, so most designs run only part of
the model inside them. DarkneTZ, for example, places selected layers in Arm
TrustZone and shows that this reduces membership-inference risk
\cite{mo2020darknetz}. The layers left outside, however, are exposed.
TEESlice shows that an attacker can exploit the public pretrained model from
which a private model was derived, undermining the assumptions behind earlier
partitioning schemes; it responds by separating public and private components
before training \cite{li2025teeslice}. The lesson is that partitions should be
chosen based on what is secret and what the attacker already knows, not on
which layers are cheapest to protect.

A related approach keeps the TEE small and offloads heavy computation to an
ordinary accelerator. Slalom verifies that an untrusted accelerator computed
linear layers correctly and can keep inputs private \cite{tramer2019slalom},
but the accelerator still sees the weights. Verifying a computation and
keeping a model secret are different goals, so a design that sends plaintext
weights out of the TEE needs a separate answer for the second.

Finally, TEEs provide no isolation against physics-based attacks. CLKSCREW and Plundervolt broke
TrustZone and SGX, respectively, by abusing the processor's own
power-management controls to induce faults
\cite{tang2017clkscrew,murdock2020plundervolt}. These are historical results
on specific implementations, but they show why a TEE cannot simply be assumed
to resist a captured-device attacker. Any such claim should state which
physical attacks were tested, which were mitigated, and which were ruled out
of scope.

\subsection{FPGA-Based Protection}

FPGAs let designers implement a model directly in custom hardware, which also
lets them build protections into that hardware. Depending on the device
family, available controls include encrypted and authenticated configuration
bitstreams, protected key storage, disabled readback and debug, on-chip
voltage and temperature sensors, and automatic key erasure when tampering is
detected \cite{amd2026tamper}. Two cautions apply. First, encrypting the
bitstream does not protect weights loaded separately from external memory;
those need their own protection. Second, tamper response cuts both ways: an
attacker who can deliberately trigger an alarm may be able to shut down a
satellite or heart monitor at a critical moment.

Configuration security is not guaranteed by the platform either. The
Starbleed attack broke both the confidentiality and the authenticity of
bitstreams on Xilinx 7-Series and Virtex-6 devices using only access to the
configuration interface \cite{ender2020unpatchable}. The flaw does not
necessarily extend to other device families, but the result shows that FPGA
security must be assessed platform by platform rather than assumed.

Even with a secure configuration, the running circuit leaks information
through power and electromagnetic emissions. Two families of countermeasures
address this. \emph{Masking} splits each sensitive value into random shares
so that no single measurement reveals it, while \emph{hiding} makes the timing
or power profile of operations less predictable. Both must cover the whole
datapath, including activation functions and memory transfers, not just the
multiplications. Both are also costly: Dubey et al.\ report a 4.7--5.4$\times$
area-delay overhead for their masked neural network designs
\cite{dubey2022guarding}, and ModuloNET explores co-designing the arithmetic
and the masking for binarized networks \cite{dubey2022modulonet}.

Shuffling, which randomizes the order in which operations execute, is a
cheaper form of hiding that has been evaluated for neural networks
\cite{brosch2022shuffling}. The randomization itself can leak, however.
BlackJack and later work on Fisher--Yates shuffling recover the shuffle order
from side-channel traces \cite{ganesan2023blackjack,puskac2025shuffling}, so
the scheduler and random-number generator need the same protection as the
model. More generally, a countermeasure should be evaluated against an
attacker who adapts to it, with its costs in area, memory, randomness,
latency, and energy reported alongside the attack results. Showing that a
naive attack fails is not enough.

\subsection{Assume Extraction: Limiting the Consequences}

Given enough resources, an attacker will likely extract something from a
captured device: an encrypted artifact, the usable weights, or a surrogate
model that mimics the original. Defenders should therefore not only resist
extraction but also limit what it is worth. The approaches below do this in
different ways. Implementation randomization slows physical extraction,
hardware binding makes a copied model useless on other hardware, diversity
limits how far an attack on one device carries to others, and fleet
monitoring detects and contains the damage.

\paragraph{Per-device model diversity.}
If every device in a fleet runs the same model, one extraction compromises
them all: an adversarial example crafted against the stolen copy works
everywhere. Deploying different models on different devices could break this
link. Varying random seeds, fine-tuning data, architectures, or ensemble
subsets all produce different models, but different weights do not guarantee
different weaknesses, since adversarial examples often transfer between
independently trained models. Prior work explicitly trains ensemble members
to differ in their gradients, predictions, or learned features
\cite{kariyappa2019diversity,pang2019diversity,yang2020dverge}. These methods
were designed to make a combined ensemble robust, however, and it remains
open whether they protect a fleet in which each device runs only one member.

We propose evaluating diversity directly at the fleet level: extract one or
several models, build attacks from them, and measure their success against
devices the attacker has not seen. Two points matter in such an evaluation.
Each device still needs adequate robustness on its own, since diversity
limits transfer but not an attack tailored to a captured unit. And the benefit
likely shrinks as the attacker captures more devices, so results should be
reported as a function of the number captured. Diversity also carries costs
beyond inference, in training, validating, and updating many models. Recent
preprints on VeilNN and boundary-aware model obfuscation pursue related goals
\cite{zhang2026veilnn,zhang2026bmo}; we emphasize evaluation under realistic
capture and deployment constraints.

\paragraph{Function-preserving transformations.}
Neural networks contain symmetries that allow the weights to be rearranged
without changing what the network computes. Consider a hidden ReLU layer
followed by an affine layer:
\begin{equation}
  h = \operatorname{ReLU}(Wx + b), \qquad z = Vh + c.
\end{equation}
For any permutation matrix \(P\) and positive diagonal matrix \(D\), the
transformed parameters
\begin{equation}
  W' = DPW, \qquad b' = DPb, \qquad V' = VP^{\mathsf T}D^{-1}
\end{equation}
give \(h' = DPh\) and therefore \(V'h' + c = Vh + c\). Reordering neurons and
scaling them by positive factors passes through the ReLU unchanged, and the
next layer exactly undoes it.

The appeal for edge security is that a device could re-randomize its weights
at every boot, or even every inference. Side-channel attacks typically
average many measurements of the same computation; if the stored and
processed values keep changing, that averaging becomes harder. This
construction, including periodic and per-inference renewal, appears in a
2024 patent application \cite{hoogerbrugge2024protection}, so our
contribution is to frame how it should be implemented and evaluated.

The limits are significant. Because the network's function is unchanged, the
transformation does nothing against adversarial examples or functional
extraction; it targets only physical leakage. Even there it is weaker than
masking, since permutation moves values without hiding them and positive
scaling preserves signs and activation patterns. Per-boot randomization still
permits many measurements within a single boot, while per-inference renewal
adds computation and storage costs. Deployment raises further issues.
Equivalence holds in exact arithmetic but may break under quantization,
overflow, and rounding; positive scaling is not a symmetry of activations
such as GELU, sigmoid, or ReLU6; and residual connections and normalization
layers must be transformed consistently. On an FPGA, randomizing memory
addresses or schedules within a fixed bitstream is likely more practical than
reconfiguring the fabric for every inference. The open question is whether
securely generated transformations raise the cost of recovering a usable
model beyond what shuffling and masking already achieve.

\paragraph{Hybrid TEE and FPGA protection.}
TEEs and FPGAs have complementary strengths, which suggests a hybrid in which
the TEE handles keys, update verification, and attestation while the FPGA
performs protected inference. The difficulty lies in the seams. The two
components must be connected by authenticated, protected links, and
attestation must cover both: RATS addresses such composite devices
\cite{birkholz2023rats}, but measuring only the CPU says nothing about the
attached accelerator. Protection must also follow the full data path. A TEE
cannot protect weights it releases onto an exposed bus, and a correctly
operating FPGA cannot prevent tampering with preprocessing before inference
or with results afterward.

Combining the hybrid with the transformations above adds a complication. If
weights are re-randomized at runtime, their hash changes constantly and
simple integrity checks fail. An integrity policy would instead need to vouch
for the approved source model and the approved transformation code, telling
legitimate re-randomization apart from model replacement without revealing
the secret transformation. Making this composition work is a research
problem, not an automatic result of pairing the two devices.

\paragraph{Hardware locking and attribution.}
Hardware-locked models function correctly only with a secret key held by the
device. HPNN is one example \cite{chakraborty2020hpnn}, but later work
recovered key information from the locked model's architecture and weights,
combined with queries returning output scores, without ever reading the key
directly \cite{li2024locking}. Keeping the key secret is therefore not
sufficient. It also helps to distinguish three outcomes: an attacker who
copies the locked model, one who recovers usable weights, and one who simply
keeps querying the device. A locking scheme may prevent the first while
leaving the other two open.

Watermarking addresses a different need: proving ownership after theft.
Uchida et al.\ embed ownership information in the weights, while Adi et al.\
train the model to respond distinctively to secret trigger inputs
\cite{uchida2017watermarks,adi2018watermarking}. A single shared watermark
identifies the vendor; tracing which device leaked requires a unique
fingerprint per device. Such marks must survive fine-tuning, pruning,
distillation, and attackers who compare multiple copies, without falsely
implicating innocent devices. Watermarking aids attribution, but it cannot
stop an adversary from using a stolen model privately.

\paragraph{Fleet integrity monitoring and recovery.}
Each device should have an approved manifest describing its model,
preprocessing, runtime, accelerator configuration, and allowed versions,
delivered through authenticated updates with rollback protection
\cite{moran2022manifest}. Attestation is only as trustworthy as its
measurement: a hash reported by an untrusted application proves little. Even
a trustworthy measurement is only a snapshot, and showing that the stored
model was correct does not prove that a later inference used it
\cite{birkholz2023rats}.

Behavioral checks complement measurement. Defenders can periodically send
canary inputs with known correct outputs and flag devices that answer
wrongly. A finite set of canaries cannot rule out an untested trigger,
however, and a compromised device may recognize canaries and answer them
correctly. Once a model has been extracted, its correct answers are no longer
secret either. Canaries matter most for a backdoor shipped through the
official release process: provenance checks will accept such a model as
approved, leaving behavioral testing as the main line of detection.

Containment and recovery complete fleet defense. Devices should hold distinct
credentials and limited authority, and no endpoint should be able to sign
releases for the fleet; model diversity cannot compensate for a shared key or
a common firmware bug. Recovery plans must address revocation,
reprovisioning, and devices with intermittent connectivity, recognizing that
revocation can cut off future updates and keys but cannot recall weights
already extracted. Finally, detection carries its own risks. Quarantining a
healthy drone or interrupting a heart monitor can be as harmful as an attack,
so policies must balance false alarms against missed compromises and define
what a disconnected device may keep doing while it cannot be verified.